\documentclass[twocolumn,nofootinbib]{revtex4}

\usepackage{amsmath,epsfig}

\begin{document}

\title{Dense-dilute factorization for a class of stochastic processes
and for high energy QCD}
\author{St\'ephane Munier}
\affiliation{Centre de Physique Th\'eorique, 
\'Ecole Polytechnique, CNRS, 91128 Palaiseau, France}

\begin{abstract}
Stochastic processes described by
evolution equations in the universality class of the
FKPP equation may be approximately factorized into
a linear stochastic part and a nonlinear deterministic part.
We prove this factorization on a model with no spatial dimensions
and we illustrate it numerically on a one-dimensional toy model
that possesses some of the main features of high energy QCD evolution. 
We explain how this procedure may be applied
to QCD amplitudes, by combining Salam's Monte-Carlo implementation of the 
dipole model and a numerical solution of the Balitsky-Kovchegov equation.
\end{abstract}
\maketitle

\section{Introduction}

High energy scattering in QCD was recently shown to 
be essentially similar to a reaction-diffusion process \cite{M2005,EGBM2005}.
To understand this correspondence, 
one needs to introduce an (unphysical) 
forward elastic scattering amplitude $T(Y,r)$,
that encodes the interaction probability
for {\it given} parton or field configurations of the incoming
hadrons, at rapidity $Y$, and for a characteristic transverse distance scale $r$.
The physical scattering amplitude
$A(Y,r)$ is the average of $T(Y,r)$ over all possible realizations
of the fields or partons.
Although $T$ is not measurable experimentally, it is a quantity
that has also to be introduced when one wants to write 
a Monte-Carlo event generator.

The equation that governs the rapidity evolution 
of $T$ at fixed impact-parameter 
belongs to the universality class of the stochastic 
Fisher-Kolmogorov-Petrovsky-Piscounov (FKPP) equation \cite{FKPP},
or equivalently, of the Reggeon field theory equation.
It can be seen as a stochastic extension of the well-known
Balitsky-Kovchegov (BK) equation \cite{B,K},
that describes a peculiar limit of high energy QCD.
It is convenient\footnote{
The reason for introducing this transformation is that the nonlinearity
present in the Balitsky-Kovchegov equation 
simplifies greatly in momentum space.} 
to go to momentum space using the transformation \cite{K}
\begin{equation}
\tilde T(Y,k)=\int \frac{d^2 r}{2\pi r^2} e^{ikr} T(Y,r).
\label{fouriertransform} 
\end{equation}
The evolution of $\tilde T$ can be written,
for example, in the form
\begin{equation}
\partial_{\bar\alpha Y}\tilde T=\chi(-\partial_{\ln k^2})\tilde
T-\tilde T^2
+\alpha_s \sqrt{2\tilde T}\nu,
\label{RFT}
\end{equation}
where $\bar\alpha=\alpha_s N_c/\pi$,
$\chi$ is a representation of the BFKL kernel \cite{BFKL}
in momentum space and $\nu$ is a noise of zero mean that varies 
randomly by typically one unit
when $\bar\alpha Y$ or $\ln k^2$ are changed by one unit.
$\tilde A$ is the average of $\tilde T$ 
with respect to the noise $\nu$.

This new approach to high energy scattering in QCD, that makes use
of ideas and tools of statistical physics, has already inspired a
number of theoretical works. Eq.~(\ref{RFT}) itself was also subsequently
discussed in Ref.~\cite{IT2004}
and studied numerically (with a Gaussian noise $\nu$) 
in Ref.~\cite{S2005}.
Many developments from different perspectives have followed:
In particular, the connection to Reggeon field theory 
was investigated \cite{KovL2006},
a QCD derivation of Eq.~(\ref{RFT}) was searched for \cite{MSW2005}, and
different formalisms to describe the same physics were proposed \cite{LL2005}.

Eq.~(\ref{RFT}) may be interpreted in the following way: It
describes the evolution 
in time $t=\bar\alpha Y$ of a fraction $\tilde T$ of $N=1/\alpha_s^2$ particles
(gluons)
per unit of $x=\ln k^2$ that multiply and
diffuse in $x$ through the branching diffusion kernel $\chi$, and that
recombine through the nonlinear term.
Eq.~(\ref{RFT}) is however not an exact equation of QCD, and should by no way 
be interpreted as such.
It is
rather a synthetic form of writing two exactly known limits:
the large parton density limit, 
in which the evolution of $\tilde T$ is given by
the Balitsky-Kovchegov equation
\begin{equation}
\partial_{\bar\alpha Y}\tilde T=\chi(-\partial_{\ln k^2})\tilde
T-\tilde T^2,
\label{BKmom}
\end{equation}
and the low density limit, represented by the linear stochastic
equation
\begin{equation}
\partial_{\bar\alpha Y}\tilde T=\chi(-\partial_{\ln k^2})\tilde
T
+\alpha_s \sqrt{2\tilde T}\nu.
\label{dipole}
\end{equation}

Let us briefly recall a bit more precisely the QCD content of Eq.~(\ref{RFT}) 
(We refer the reader to the original papers \cite{M2005,EGBM2005}
for details). In the low density 
region of phase space (equivalently, in the region 
in which the amplitude is small, $T,\tilde T\ll 1$), 
it is useful to view rapidity evolution in the framework of the  
color dipole model \cite{M1993}. 
One goes to the reference
frame of one of the interacting hadrons, 
that we assume to be a quark-antiquark dipole for simplicity and 
that we will call the probe.
The second hadron carries all the rapidity and QCD evolution
and will be referred to as the target.
In the large-$N_c$ limit, 
the Fock state of the target may be represented
by a set of color dipoles.
Each of these dipoles, characterized by a two-dimensional 
position vector $x_{01}=x_0-x_1$
in transverse coordinate space, 
is spanned by two gluons respectively sitting at
positions $x_0$ and $x_1$.
The model is defined
by the stochastic branching of the dipoles that occurs as the target is boosted 
to larger rapidities:
In a step $dY$ of rapidity, each dipole present
at rapidity $Y$
has the probability 
\begin{equation}
{\bar\alpha} dY \frac{d^2 x_2}{2\pi} \frac{x_{01}^2}{x_{02}^2 x_{12}^2}
\label{dipoles}
\end{equation}
to be replaced by two new dipoles of respective sizes $x_{02}$ and $x_{12}$.

When nonlinear effects can be neglected,
the relationship between 
the number of dipoles $n(Y,x_{01})$ 
and the amplitude $T(Y,r_{01})$ 
for the scattering of the probe dipole of size $r_{01}$ off 
a random configuration of the target
reads, in appropriate normalizations (see e.g. \cite{IM2003}),
\begin{multline}
T(Y,r_{01})=\frac{\pi^2\alpha_s^2}{2}\\
\times\int \frac{d^2 x_0}{2\pi} \frac{d^2 x_1}{2\pi}
\ln^2\frac{|r_0-x_1|^2|r_1-x_0|^2}{|r_0-x_0|^2|r_1-x_1|^2}n(Y,x_{01}).
\label{corresp}
\end{multline}
This relationship turns out to be
approximately local
in dipole sizes and impact parameter:
$T(Y,r_{01})$ is of order $\alpha_s^2$ times the number of dipoles
in the considered configuration of the target whose sizes are
in a bin of width 1 (on a logarithmic scale) centered on $|r_{01}|$, and
which sit within a distance $|r_{01}|$ of the probe in impact
parameter.

By converting the splitting process~(\ref{dipoles}) 
into an evolution equation for
$T$ with the help of Eq.~(\ref{corresp}), by 
transforming to momentum space using Eq.~(\ref{fouriertransform}),
and by averaging over the angle
in the transverse plane, one would get 
the linear terms in Eq.~(\ref{RFT})
together with an appropriate noise $\nu$ 
that would encode the fluctuations in the
dipole number induced by
the stochastic branching process~(\ref{dipoles}): this is
precisely Eq.~(\ref{dipole}).

Dipole branching
leads to an exponential increase of
their number with rapidity, and consequently,
to an unlimited rise of $T$ if formula~(\ref{corresp}) is applied: 
This would be
inconsistent with unitarity, which requires the bound $T\leq 1$.
It is believed that this rise is tamed
by nonlinear effects \cite{GLRMQ} that limit the number
of dipoles in the target. Unfortunately,
 the latter effects have not yet been formulated in the
framework of the dipole model. 
It is not even clear that dipoles should
still be the relevant degrees of freedom in that regime.

On the other hand, one knows the equation 
that gives the average
scattering amplitude at rapidity $Y+dY$ given the amplitude at rapidity $Y$, 
with 
the exact nonlinearity
that preserves the unitarity of $T$. It reads:
\begin{multline}
\langle T(Y+dY,r_{01})| T(Y,r_{01})\rangle=
T(Y,r_{01})+\\
\bar\alpha dY\int \frac{d^2 r_2}{2\pi}\frac{r_{01}^2}{r_{02}^2 r_{12}^2}
(T(Y,r_{02})+T(Y,r_{12})-T(Y,r_{01})\\
-T(Y,r_{02})T(Y,r_{12})).
\label{BK}
\end{multline}
This equation is equivalent to
the first equation in the celebrated Balitsky hierarchy \cite{B}.
Eq.~(\ref{BK}) may be obtained from Eq.~(\ref{RFT}) 
by averaging it over the noise $\nu$ 
between rapidities $Y$ and $Y+dY$. 
A Fourier transformation to position space
completes the identification.
In practice, Eq.~(\ref{BK}) is useful in a regime in which $T$ may be
approximated by its average, i.e. when $\langle T(Y+dY,r_{01})| T(Y,r_{01})\rangle$
may be replaced by $T(Y+dY,r_{01})$, turning~(\ref{BK}) into a closed equation
for $T=\langle T\rangle$, Eq.~(\ref{BKmom}).
This is realized when the underlying effective number of dipoles
is large, i.e. in the region of high density.
The corresponding equation is the BK equation
\cite{B,K}
and is in the universality class of the (deterministic) FKPP equation
\cite{MP2003}.  
The evolution of higher order correlators of $T$s 
that would be needed to be able to evolve $T$ also
outside the dense region
has still not been derived in a systematic way for the generic
case of dipole-dipole scattering. 

Although the correct formalism to describe the evolution of $T$
accurately in all regimes is still not available,
exact asymptotic results~\cite{BDMM2005}
universal enough to also apply to QCD amplitudes
have been found from the approximate 
formulation~(\ref{RFT}). Unfortunately, their validity is quite
reduced since the
limit $\ln(1/\alpha_s^2)\gg 1$ had to be assumed, 
which is of course out of experimental reach.
Hence it would be useful to understand what can be expected
quantitatively from what is known up to now,
beyond the far asymptotics.

The goal of this paper is to
show that quite accurate results for the evolution of the QCD scattering
amplitudes may be extracted
from a numerical study  that appropriately matches
a Monte-Carlo simulation of the color dipole model, valid in the dilute
regime, to the
evolution of the amplitude given in 
Eq.~(\ref{BK}), useful in the dense regime.
We shall propose a factorization
procedure of $T$, on an event-by-event basis, 
that we justify by a calculation in the framework of a zero-dimensional 
model and motivate and test on a
one-dimensional toy model. 
We then explain how this factorization could be applied to QCD.


\section{Factorization in a zero-dimensional model}

In a first stage, we study a zero-dimensional model
(transverse variables are not considered)
in order to introduce the factorization rigorously.
Zero-dimensional models were investigated
for possible applications to QCD
some time ago \cite{M1994,Salam}.
Solutions for models of that kind have recently been 
discussed in Refs.~\cite{SX2005,KL2006}, from different
perspectives. The solutions which 
have been obtained will help us to assess
the validity of the factorization that we shall 
propose here.

\subsection{Definition of the model}

We investigate  the time evolution of
a specific Markovian process in the universality class of the 
zero-dimensional stochastic FKPP equation.
We consider a system of $n(t)$ particles. Between times $t$ and $t+dt$,
each particle has a probability $p_+=dt$
to split in two particles.
For each pair of particles,
there is a probability $dt/N$ that one of
them is lost, 
and thus each given particle has probability $p_-=(n_t-1)dt/N$
to disappear.
These rules completely define the process.

There are several ways to represent this evolution.
We may write the distribution of the number $n_{t+dt}$
of particles at time $t+dt$
given the number of particles $n_t$ at time $t$:
\begin{multline}
P(n_{t+dt}|n_t)=\delta_{n_{t+dt},n_t}
\left(1-n_t dt -\frac{n_t(n_t-1)}{N}dt\right)\\
+\delta_{n_{t+dt},n_t+1} n_t dt
+\delta_{n_{t+dt},n_t-1} \frac{n_t(n_t-1)}{N} dt.
\label{probazero}
\end{multline}
This equation may be cast in the form of a stochastic 
equation for $n_t$
by first computing the mean and variance of $n_{t+dt}$ given
$n_t$, with the help of Eq.~(\ref{probazero}). 
This enables one to write the time evolution
of $n_t$ in terms of a drift and of a noise of zero-mean
and normalized variance, namely:
\begin{equation}
\frac{dn_t}{dt}=n_t-\frac{n_t(n_t-1)}{N}+
\sqrt{n_t+\frac{n_t(n_t-1)}{N}}\nu_{t+dt},
\label{evoln}
\end{equation}
where $\nu$ is such that $\langle \nu_t\rangle=0$ and 
$\langle\nu_t\nu_{t^\prime}\rangle=\delta(t-t^\prime)$.
Note that the distribution of $\nu$ 
depends on $n_t$ and
is not a Gaussian.
This last point is easy to understand: The evolution
of $\nu_t$ is intrinsically discontinuous, since
it stems from a rescaling of
$n_t$, which is an integer at all times.
A Brownian evolution (i.e. with a Gaussian noise) would
necessarily be continuous.
For completeness, we write the statistics of  $\nu_{t+dt}$, which is
easy to derive from the evolution of $n$:
\begin{equation}
\nu_{t+dt}=
\begin{cases}
\phantom{-}\frac{1}{\sigma\, dt}-\frac{\Delta}{\sigma} & \text{proba}\
{n_t\, dt}\\
\phantom{-\frac{1}{\sigma\, dt}}-\frac{\Delta}{\sigma} & \text{proba}\
1-n_t\, dt-\frac{n_t(n_t-1)}{N}dt
\\
-\frac{1}{\sigma\,dt}-\frac{\Delta}{\sigma} & \text{proba}\
\frac{n_t(n_t-1)}{N}dt,
\end{cases}
\end{equation}
where
$\Delta=n_t-\frac{n_t(n_t-1)}{N}$ and
$\sigma=\sqrt{n_t+\frac{n_t(n_t-1)}{N}}$.
We see well the jumps induced by the terms proportional to $1/dt$.

Formulating the process with the help of a stochastic equation such
as~(\ref{evoln}) has no real advantage at this point.
Here, we just aimed at showing explicitely the connection with
stochastic partial differential equations.


\subsection{Poissonian states and ``Pomerons''}

Instead of looking at a state of definite occupancy,
one could also consider the evolution of a Poissonian state
whose occupation numbers are distributed as
\begin{equation}
P_{z_t}(n_t)=\frac{z_t^{n_t}}{n_t!}e^{-z_t}
\label{poisson}
\end{equation}
and follow the evolution of $z_t$, that is to say, compute
the probability distribution of $z_{t+dt}$ given $z_t$, 
$P(z_{t+dt}|z_t)$.
As can be easily checked,
the moments of $z_{t+dt}$
are the factorial moments of $n_{t+dt}$.
This statement may be written as
\begin{multline}
\sum_{n_t,n_{t+dt}}\frac{n_{t+dt}!}
{(n_{t+dt}-k)!}P(n_{t+dt}|n_t)P_{z_t}(n_t)\\
=\int dz_{t+dt}
z_{t+dt}^k
P(z_{t+dt}|z_t).
\label{factorial}
\end{multline}
Replacing Eqs.~(\ref{probazero}) and~(\ref{poisson}) in Eq.~(\ref{factorial}), 
one finds
\begin{multline}
\int dz_{t+dt}
z_{t+dt}^k
P(z_{t+dt}|z_t)
=z_t^k+dt\left[z_t-\frac{z_t^2}{N}\right]k z_t^{k-1}\\
+\frac12 \left[{2dt}\left(z_t-\frac{z_t^2}{N}\right)
\right]k(k-1)z_t^{k-2},
\end{multline}
from which it is obvious that $P(z_{t+dt}|z_t)$ is
a Gaussian of mean $z_t+dt(z_t-\frac{z_t^2}{N})$
and variance ${2}{dt}{(z_t-\frac{z_t^2}{N})}$.
In the same way as one translates Eq.~(\ref{probazero}) 
into Eq.~(\ref{evoln}), this may be
expressed in the form of a stochastic evolution equation for $z_t$ \cite{PL}
\begin{equation}
\frac{dz_t}{dt}=z_t-\frac{z_t^2}{N}+
\sqrt{2\left(z_t-\frac{z_t^2}{N}\right)}\eta_{t+dt}.
\label{itosfkpp}
\end{equation}
This is an Ito equation:
$\eta$ is a Gaussian white noise satisfying $\langle \eta_t\rangle=0$ and
$\langle \eta_t\eta_{t^\prime}\rangle=\delta(t-t^\prime)$.
$z_t$ is now evolving continuously, unlike $n_t$.
Eq.~(\ref{itosfkpp}) 
is the zero-dimensional version of the stochastic FKPP equation.
It is of course consistent with Eq.~(\ref{evoln}), since the
moments of $z$ are the factorial moments of $n$.
Starting from Eq.~(\ref{itosfkpp}) and transforming it into a hierarchy
for the factorial moments of $n_t$,
one can compute the first few orders of $\langle n_t\rangle$
in a $1/N$ expansion resummed for large $t$ \cite{SX2005}.
The result turns out to be an asymptotic Borel-summable series.
It reads \cite{M1994,SX2005}
\begin{equation}
\langle n_t\rangle
=N\sum_{k=1}^\infty (-1)^{k+1}{N}^{-k} k! e^{k t}.
\label{borelseries}
\end{equation}
Each term may be interpreted as the result of the evaluation
of a diagram with a number $k$ of ``Pomerons'' being exchanged in the $t$-channel.
We refer the reader to \cite{SX2005} for a detailed calculation in such
a framework.

The series~(\ref{borelseries}) can 
be rewritten in the form of a Borel integral,
\begin{equation}
\langle n_t\rangle
=N\left(
1-Ne^{-t}\int^{\infty}_{0}
\frac{db}{1+b}e^{-N e^{-t}b}
\right),
\label{borel}
\end{equation}
which eventually may be expressed in terms of special functions.

Eqs.~(\ref{borelseries}) and ~(\ref{borel}) are
 a priori valid for $e ^t/N\ll 1$.
Actually, it was found in Ref.~\cite{SX2005} 
from the exact evaluation of subleading orders,
that its effective
range of validity is much broader, $t/N\ll 1$.

The classical limit is also easily identified from Eq.~(\ref{itosfkpp}):
Indeed, a classical state has a definite value of $z$.
The classical path 
$z(t)=\langle n_t\rangle\simeq n_t^\text{MF}$ is obtained 
by solving the deterministic 
part of Eq.~(\ref{itosfkpp}), namely
\begin{equation}
\frac{dn^\text{MF}_t}{dt}=n^\text{MF}_t-\frac{(n^\text{MF}_t)^2}{N}.
\label{MF}
\end{equation}
This approximation
is expected to be valid only when
the typical number of particles in the system is large.
It is useful to notice that the steady fixed point of the evolution
is $\langle n_t\rangle=N$.

The formulation of Eq.~(\ref{itosfkpp}) is
helpful for analytical calculations of the moments of $n_t$. 
However, in order to describe a physical system
that starts evolution with a definite number of particles, one
needs to introduce complex values of the Poisson parameter $z$.
This makes Eq.~(\ref{itosfkpp}) of little interest for numerical simulations
of such systems.
Furthermore, the analytical method for the computation of moments
is very awkward to transpose
numerically, 
since it leads to an asymptotic series
that eventually needs to be resummed.
Finally, it is not straightforward to generalize to models with
spatial dimensions.

\begin{figure}
\epsfig{file=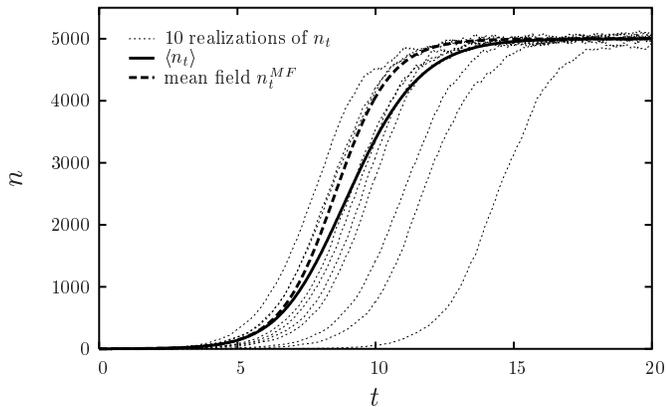,width=9cm}
\caption{\label{fig1}
{\em Bunch of dotted lines:}
Ten different paths for $n_t$ obtained from
the evolution~(\ref{evoln}) of the 
zero-dimensional model starting from a single-particle initial
condition.
{\em Dashed line:}
Mean-field solution, see Eqs.~(\ref{MF}),(\ref{resMF}). 
{\em Full line:} Numerical solution
for $\langle n_t\rangle$ obtained by averaging over a large
number of realizations of Eq.~(\ref{evoln}).
}
\end{figure}


\subsection{Solving the stochastic evolution for $n$:
dense-dilute factorization}

For the difficulties mentioned above with the Poissonian state
approach, 
we wish to take a different point of view on the evolution of
the stochastic model. Instead of performing 
a large-$N$ perturbative calculation directly of 
$\langle n_t\rangle$ using field-theoretical methods, 
we try to characterize 
the shape of each realization of an evolution that starts with $n_0=1$
particles. 
What we mean by ``realization'' is a given path for $n$
generated by the stochastic evolution~(\ref{evoln}).

It is useful to visualize a few such
realizations: This is shown in Fig.~\ref{fig1} for $N=5000$,
together with the solution to the mean field equation~(\ref{MF}).
One sees that the curves that represent $n_t$ 
look like the solution to the mean-field equation~(\ref{MF}),
but with the origin of times translated by some random $t_0$.
(The curves look also slightly noisy around the average trend,
but the noise would be still much weaker for larger values of $N$).
This suggests that once there are ``enough'' particles in the system, 
say $\bar{n}\gg 1$, the evolution becomes essentially deterministic.
Hence stochasticity only manifests itself in the initial stages of the
evolution until $n_t=\bar n$, 
but in a crucial way: Indeed, as seen in Fig.~\ref{fig1}, after averaging,
$\langle n_t\rangle$ is significantly different from $n_t^\text{MF}$, and this
difference stems from rare realizations in which the particle number stays low
for a long time.
Therefore, in individual realizations,
stochasticity should be taken into
account exactly as long as $n_t<\bar n$. 
Fortunately, when the number of particles in the system is small
compared to the maximum number of particles $N$, the stochastic evolution
is essentially governed by a linear equation which is not 
difficult to handle analytically.

This heuristical discussion suggests that we may factorize 
the evolution in a linear stochastic
evolution up to the time at which the number $n$ of particles in the system
reaches $\bar{n}$, and continue
through a nonlinear but deterministic equation, which is obtained from
a mean field
approximation to the evolution equation.
As we will see, this simple observation leads to an elegant computation of
$\langle n_t\rangle$ that consistently agrees with the lowest order 
(see Eq.~(\ref{borel}))
derived in Ref.~\cite{M1994,SX2005}.

Let us denote by $p_{\bar{n}}(\bar{t})$ the 
distribution of the times at which the number of particles in the system
reaches $\bar{n}$ and $\langle n_t|n_{\bar t}\rangle$ the conditional average
number of particles at time $t$ given that there are $n_{\bar t}$ particles
in the system at time $\bar t$.
One may write
\begin{equation}
\langle n_t\rangle=
\int_0^\infty d\bar{t}\, p_{\bar{n}}(\bar{t})
\langle n_{t}|n_{\bar{t}}\rangle.
\label{procedure0}
\end{equation}
So far, this expression is exact.

We now assume that the evolution is linear when $n_t <\bar n$
and deterministic for $n_t>\bar n$. Thus, in the previous equation, we replace 
$p_{\bar{n}}(\bar{t})$ by the solution
$p_{\bar{n}}^{\text{lin}}(\bar{t})$
of the linear problem obtained by setting
$p_+=dt$ and $p_-=0$.
Furthermore, we approximate $\langle n_{t}|n_{\bar{t}}\rangle$
by the solution to the nonlinear evolution in the mean
field approximation~(\ref{MF}) over a time interval $t-\bar t$
starting with $\bar n$ particles at time $\bar t$, 
that we denote by
$n^\text{MF}_{t-\bar t|\bar n}$ .
In these notations, we find the factorization
\begin{equation}
\langle n_t\rangle=
\int_0^\infty d\bar{t}\, p^\text{lin}_{\bar{n}}(\bar{t})
n^\text{MF}_{t-\bar{t}|\bar{n}}.
\label{procedure}
\end{equation}
Note that for $\bar t>t$ (i.e. when there are less than $\bar n$ particles
in the system at the considered time $t$), 
$n^\text{MF}_{t-\bar{t}|\bar{n}}$ is like
a backward evolution towards lower number of particles.
 This is not a problem since we then go back to the
dilute regime, and the solution for $\langle n\rangle$ in that regime
is just obtained by taking the dilute limit of $n^\text{MF}$, given by
the solution of the equation obtained from Eq.~(\ref{MF}) by dropping
the nonlinear term.

We claim that this factorization is valid whenever
$\bar n$ is large enough compared to 1 to justify the mean field 
approximation 
for the subsequent evolution, but, at the same time, $\bar n$ 
is small compared to $N$ 
in such a way that the evolution up to $\bar n$ be linear.

Let us express give explicit expressions for the different quantities
that appear in Eq.~(\ref{procedure}).
The solution to Eq.~(\ref{MF}) reads
\begin{equation}
n^\text{MF}_{t-\bar{t}|\bar{n}}=\frac{N}{1+\frac{N}{\bar{n}}e^{-(t-\bar{t})}}
\label{resMF}
\end{equation}
for $\bar n \ll N$.
It is also not difficult to show that $p_{\bar{n}}^\text{lin}$
solves the equation
\begin{equation}
p^\text{lin}_{\bar{n}}(\bar t)
=(\bar{n}-1)\int_0^{\bar t} d{\bar t}^\prime\, p^\text{lin}_{\bar{n}-1}({\bar t}^\prime)
e^{-(\bar{n}-1)({\bar t}-{\bar t}^\prime)}.
\label{evolpn}
\end{equation}
Its solution may be found by standard generating function methods
and takes the simple form
\begin{equation}
p^\text{lin}_{\bar n}({\bar t})
=(\bar n-1)e^{-{\bar t}}(1-e^{-{\bar t}})^{\bar n-2}.
\end{equation}
In the limit of interest, that is
for large $\bar{n}$ and $\bar{t}$, it boils down to a Gumbel
distribution
\begin{equation}
 p^\text{lin}_{\bar n}({\bar t})=\bar n e^{-{\bar t}-\bar n e^{-{\bar t}}}.
\label{reslin}
\end{equation}
Replacing Eqs.~(\ref{resMF}) and~(\ref{reslin}) in Eq.~(\ref{procedure}),
we find
\begin{equation}
\langle n_t\rangle=N \int_0^\infty d\bar{t}\,
\frac{\bar{n} e^{-\bar{t}-\bar{n} e^{-\bar{t}}}}
{1+\frac{N}{\bar{n}}e^{-(t-\bar{t})}}.
\end{equation}
Since the Gumbel distribution is strongly damped for $\bar t<0$
(by a factor $e^{-\bar n}$), we may safely replace the lower limit of the integral
by $-\infty$.
Finally, the change of variable $b=\frac{\bar{n}}{N}e^{t-\bar{t}}$ 
brings the integral in the form (\ref{borel}).
The result is independent of $\bar n$, as it should be, but in numerical applications,
 one should
keep in mind that the limits $1\ll \bar n\ll N$
have been assumed, and thus finite $\bar n$ corrections must be expected.

In this picture, the Borel integral~(\ref{borel}) has 
a transparent interpretation:
the parameter $b$ is related to the factorization time $\bar t$, the
exponential is the distribution of $\bar t$ and the denominator
corresponds to the mean-field part. Our method allowed to directly arrive
at the leading order result found in Refs.~\cite{M1994,SX2005}, 
without having to resum an asymptotic (divergent) series.


\section{A one-dimensional toy model}

We shall now formulate and test this dense-dilute factorization 
on a one-dimensional toy model.
Again, we consider a system made of typically $N$ particles in its
steady state. 

With one space dimension labeled by the real variable $x$,
for large enough times, the particles form
a front that travels towards say larger values of $x$ when time
flows \cite{PVS, EGBM2005}. 
Its position $X_t$ may be taken as the position of the
$(N/2)$-th rightmost particle.
This front connects a dilute region, to the right, to a denser
region to the left. There is a point in the front 
around which the typical number of objects
is $\bar n$.
In the spirit of the previous section,
we will treat the evolution to the right of this point as stochastic
and solve a mean field equation to the left.
We will explain the factorization and illustrate it numerically 
on a specific particle model that was introduced 
in Ref.~\cite{BDMM2006}.

When one takes a new step in time $t\rightarrow t+\Delta t$, the particle at
position $x_i(t)$ is replaced, with probability $\Delta t$, by two particles at
positions, where
$\delta_1$ and $\delta_2$ are distributed according to 
$\psi(\delta_1)d\delta_1$ (resp. $\psi(\delta_2)d\delta_2$).
This rule defines the evolution of the number of particles
in a way analogous to the dipole splitting rule in QCD.
To implement a simple form of saturation of the number of 
particles,
only the $N$ particles which have largest positions are kept for 
subsequent evolution.

We define $T(t,x)$ as the fraction of these $N$ particles 
that have positions
larger than $x$ at time $t$, that is
\begin{equation}
T(t,x)=\frac{1}{N}\sum_{i=1}^N \Theta(x_i(t)-x).
\label{T-n}
\end{equation}
Obviously, $T(t,-\infty)=1$ and $T(t,+\infty)=0$ for $t$ large enough
for the total number of particles in the system to have reached $N$.

Considering for a
while the $N=\infty$ limit,
the mean evolution of $T$ in one elementary step in time reads
\begin{multline} 
\langle T(t\!+\!\Delta t,x)|T(t,x)\rangle\\
=\min\left(
1,(1-\Delta t)T(t,x)+2\Delta t \int d\delta\,\psi(\delta)T(t,x-\delta)
\right).
\label{evolT}
\end{multline}
This equation is the analogous of 
the first equation in the Balitsky hierarchy
in the QCD case, see Eq.~(\ref{BK}).
In the infinite $N$ limit, 
it is known that the large time solutions are traveling waves 
whose average velocity is determined by the linearized part of 
Eq.~(\ref{evolT}). If one defines
\begin{equation}
v(\gamma)=\frac{1}{\Delta t}
\frac{1}{\gamma}\ln\left(1-\Delta t+2\Delta t\int d\delta\, e^{\gamma\delta} \psi(\delta)
\right),
\label{velocity}
\end{equation}
then $v(\gamma)$ admits a minimum at $\gamma=\gamma_0$ and
the large-time front velocity is $v(\gamma_0)$. 
At finite $N$, the position of the front becomes stochastic,
with nontrivial (non-Gaussian) statistics.
The moments of the position of
the front are known for large $N$ \cite{BDMM2005}. The first two of them read 
\begin{subequations}
\begin{align}
V\times t\equiv&\langle X_t\rangle=v(\gamma_0)t-
\frac{\pi ^2\gamma_0 ^2 v ^{\prime\prime}(\gamma_0)}
{2\ln ^2 N}t,
\label{theoryV}
\\
D\times t\equiv&\langle X_t^2\rangle-\langle X_t\rangle^2=
\frac{\pi ^4\gamma_0 v ^{\prime\prime}(\gamma_0)}{3\ln ^3 N}t.
\label{theoryD}
\end{align}
\label{theory}
\end{subequations}
The average position of the front
was first obtained by considering a deterministic 
evolution equation with a cutoff \cite{BD,MS2004}
that simulates the discreteness of the particles in the system \cite{BD}. 
In our case, one would write
\begin{multline} 
T^\text{cutoff}(t\!+\!\Delta t,x)
=\min\bigg(
1,
(1-\Delta t)T^\text{cutoff}(t,x)\\
+2\Delta t \int d\delta\,\psi(\delta)T^\text{cutoff}(t,x-\delta)
\bigg)\\
\times
\Theta[T^\text{cutoff}(t+\Delta t,x)-{{1}/{N}}].
\label{detcutoff}
\end{multline}
The first correction to this approximation 
(not shown in Eq.~(\ref{theoryV}))
is also known \cite{BDMM2005}.
It is due to particles that are randomly sent ahead of the deterministic
part of the front (to the left of the cutoff), 
at some distance to the right of the tip of the front.
Their multiplication through time evolution
pulls the front forward, and generates a 
positive correction to the velocity and the dispersion in the
front position $D$ given in Eq.~(\ref{theoryD}).
One also knows that the system is completely renewed every
$t\sim 1/D$ units of time \cite{BDMM2006}. This remark will help
the analysis of the numerical data.

In our numerical implementation, we choose
$\psi$ to be the uniform distribution in the interval $[0,1]$, i.e.
\begin{equation}
\psi(\delta)=\Theta(\delta)\Theta(1-\delta),
\label{psi}
\end{equation}
and $\Delta t=0.1$.
Solving $v^\prime(\gamma_0)=0$ where $v(\gamma)$ is given by
Eq.~(\ref{velocity}) with these settings, we get
\begin{equation}
\begin{split}
\gamma_0&=1.46256\cdots, \\
v(\gamma_0)&=2.07006\cdots, \\
v^{\prime\prime}(\gamma_0)&=0.753472\cdots.
\end{split}
\end{equation}
$\gamma_0$ is the logarithmic slope of the falloff of the front.
$v(\gamma_0)$ is the velocity in the $N=\infty$ limit. $v^{\prime\prime}(\gamma_0)$
is a parameter that appears when one considers finite-$N$ corrections.

We first solve the complete stochastic model.
We take an initial condition of the form $x_1=\cdots=x_N=0$ (that is
$T(t=0,x)=\Theta(-x)$) and
evolve it in time using the exact evolution rules.
We evolve the system over $p\times 500$ units in $t$ ($p$ is 500 in our simulation), 
recording the
position of the front $X_{t_i}$ every $\delta t=t_i-t_{i-1}=500$ 
units of time.
We compute $V=\langle \delta_i X_{t}\rangle/\delta t$,
where $\delta_i X_{t}=X_{t_i}-X_{t_{i-1}}$,
and $D=(\langle \delta_i X_t^2\rangle -\langle \delta_i X_t\rangle ^2)/\delta t$. 
The average is over the $p$ periods of $\delta t$ units of time in one single
realization, but
for an ergodic system, averaging over the time evolution of one realization 
is like averaging over
many independent realizations of the evolution.

\begin{figure}
\epsfig{file=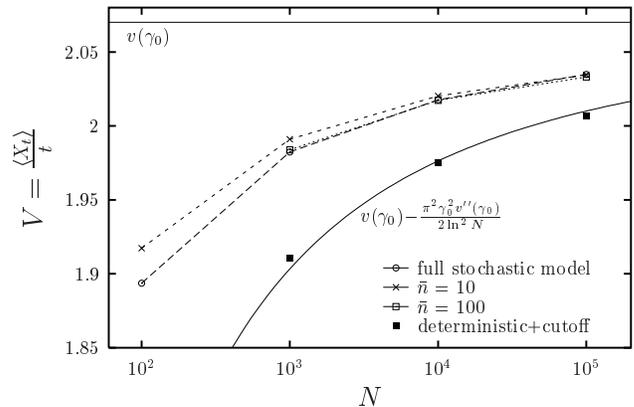,width=9cm}
\caption{\label{fig2}
Average velocity of the front $V$ versus the equilibrium
number of particles $N$ for the toy model.
The full curves represent theoretical estimates. The black squares
is the result of the mean-field calculation with a cutoff.
The solution of the full stochastic model is shown with circles
linked by large dashed lines. The solution of the mixed method for
$\bar n=10$ is displayed with crosses linked by short dashed lines.
White squares denote the case $\bar n=100$, but they are almost
indistinguishable from the solution of the full stochastic model.
}
\end{figure}

\begin{figure}
\epsfig{file=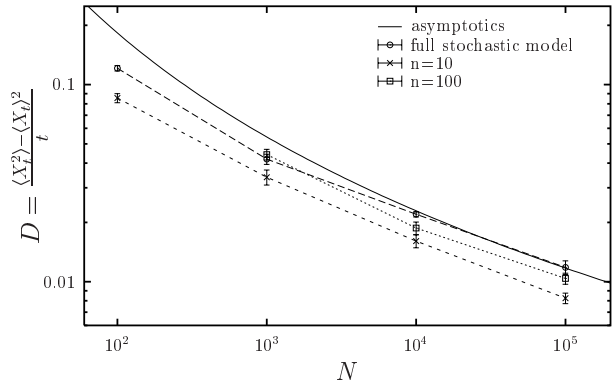,width=9cm}
\caption{\label{fig3}
Diffusion coefficient of the front $D$ versus $N$.
Same legend as in Fig.~\ref{fig2}.
}
\end{figure}

Since the system renews itself every $1/D$ units of time,
and since $1/D$ is at most of order 100, the differences of 
successive positions $\delta_i X_t$ 
are essentially independent random
variables. Hence we can estimate the statistical uncertainties by
splitting our set of positions $\delta_i X_t$ in say 10 subsets, and by evaluating
the dispersion of $V$ and $D$
between these different subsets. This helps us to provide error bars.
The results are shown in Tab.~\ref{tab1} and in Figs.~\ref{fig2} and~\ref{fig3} 
for $N=10^2,10^3,10^4,10^5$.

We also solve the deterministic evolution with a cutoff,
defined in Eq.~(\ref{detcutoff}). To this aim,
we need to take a discretization in the $x$ variable: We define a
lattice with 1000 points per
unit of $x$. The integration over $\delta$ in Eq.~(\ref{detcutoff}) 
is performed using the
rectangle method. Although this method converges very slowly
(one expects a systematic error of the order of $0.1\%$ in our settings), it is
quite well suited here because as we impose a cutoff at each step of the
evolution, $T^\text{cutoff}$ has sharp discontinuities.
The result is shown in  Fig.~\ref{fig2} together with the theoretical
formula~(\ref{theoryV}), with which there is a very good agreement
except for the lowest values of $N$. 
But that is expected since Eq.~(\ref{theoryV}) was
obtained in a large-$N$ limit of the deterministic evolution
with a cutoff.
The slight discrepancy (of order $0.1\%$) visible at large $N$
is consistently explained by the fact that our discretization
and our integration method amount to solving a model on
a lattice rather than the original model in the continuum, 
for which one can compute the asymptotic 
front velocity $v(\gamma_0)$, which is indeed slightly lower.

\begin{figure}
\epsfig{file=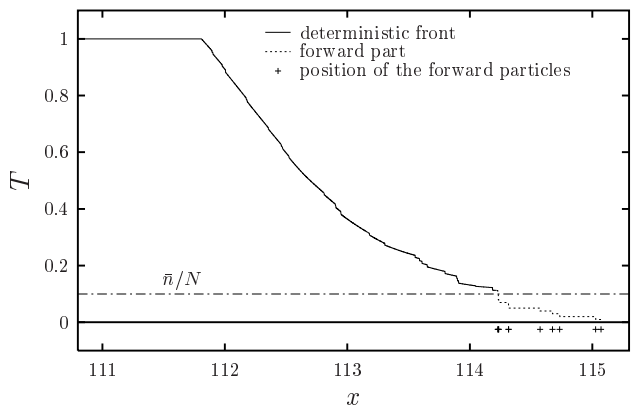,width=9cm}
\caption{\label{fig4}One realization of the front for $N=100$
evolved using the mixed method with a factorization scale $\bar n=10$.
Below $x_\text{m}$, the front is represented by $T$ evolved deterministically.
For $x>x_\text{m}$, the position of each particle is recorded and evolved
using the exact stochastic rules.
}
\end{figure}

\begin{figure}
\epsfig{file=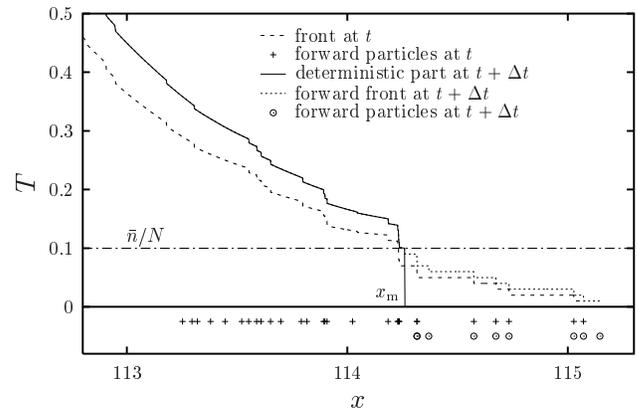,width=9cm}
\caption{\label{fig5}
One realization of the evolution of the front of Fig.~\ref{fig4} 
(reproduced here in dashes) over
the time interval $\Delta t$. The backward part is
evolved through Eq.~(\ref{evolT})
(full line). The crosses mark the positions 
of the particles that are used as an
input for the stochastic evolution of the forward tail
(they include the particles represented in Fig.~\ref{fig4} whose
positions are tracked exactly, and additional particles generated from
$T$).
The circles denote the particles that are kept at time $t+\Delta t$:
most of them were already present at $t$, and two of them 
have been produced stochastically. The forward part of 
$T$ at time $t+\Delta
t$ is represented in short dashes.
}
\end{figure}

Next, we perform a mixed evolution, applying our dense-dilute 
factorization procedure.
This goes as follows.
At all times, the forward part of the front where $T(t,x)\leq\bar n/N$ is
represented by the positions of the
$\bar n$ foremost particles, and the latter are
evolved using the exact rule (we must drop the nonlinearity, but for
this model 
it only amounts to selecting the $N$ most forward particles: 
If $\bar n$ is small enough,
the forward particles alone cannot have more than $N$ offsprings in one iteration).
This is illustrated in Fig.~\ref{fig4} for a particular realization and for
a particular choice of the parameters $N$ and $\bar n$.
For  $T(t,x)>\bar n/N$, we directly evolve $T$ using
Eq.~(\ref{evolT})
with the approximation 
$\langle T(t\!+\!\Delta t,x)|T(t,x)\rangle=T(t\!+\!\Delta t,x)$,
forgetting about the exact position of each individual particle.
More technically, an evolution over the time interval $\Delta t$ goes as follows.
We first generate particle positions from $T$
 in the region of $x$ between $x_{\bar n}-1$
and $x_{\bar n}$. Indeed, in our model, it is precisely 
the particles present in that interval 
that may split into the forward
region where we decided to keep the full stochasticity.
This is related to our particular choice of $\psi$, see Eq.~(\ref{psi}),
but in practice, it is enough that the splittings be local enough in $x$.
This is indeed the case for the fixed impact parameter BFKL evolution 
in QCD (see Eq.~(\ref{RFT}) with $x=\ln k^2$).
To get back the particle positions from the profile of $T$ 
is extremely straightforward in this particular model: 
the position $x_i$ of particle
number $i$ is the rightmost point for which $T(t,x_i)=i/N$ holds
(In more subtle models such as QCD, one would have to invert a
relation like Eq.~(\ref{corresp})).
The positions of the obtained particles
are shown with crosses in Fig.~\ref{fig5}.
We then evolve these additional particles together with the $\bar n$ 
forward particles using the full stochastic rule.
Simultaneously, we evolve $T$ deterministically for all $x$, but
we cut the result at $x_\text{m}$ defined as the position in
the front for which $T(t+\Delta t,x_\text{m})=\bar n/N$. 
Indeed, only for $T>\bar n/N$ do we trust the mean-field 
approximation
$\langle T(t\!+\!\Delta t,x)|T(t,x)\rangle=T(t\!+\!\Delta t,x)$.
In turn, we only keep the
forward particles that have positions larger than $x_\text{m}$.
Finally, the matching of the forward and backward parts of the front
is done by requiring 
that $T$ be a decreasing function
of $x$. In the case in which 
there is a number of particles larger than $\bar n$ produced
ahead of $x_\text{m}$, then $T$ 
computed from the forward particles is
larger than $\bar n/N$ at $x_\text{m}$:
We choose to continue $T(t+\Delta t,x_\text{m})$
for $x<x_\text{m}$ until the point at which 
$T(t+\Delta t,x)=T(t+\Delta t,x_\text{m})$.
Note that with this prescription for the matching, the particle distribution
in the forward part of the front is exact at all times, 
but the number of
particles in the backward part is slightly overestimated on the average.
Other prescriptions may have been chosen: They would lead to similar
quantitative results.
One particular realization of the resulting front 
is shown in Fig.~\ref{fig5}.

We turn to the discussion of the numerical results for the
mixed method.
Recall that the
factorization scale $\bar n$ has to be at the same time 
larger than 1 in such a way that mean field
evolution starting from $\bar n$ can be justified, and much smaller
than $N$ so that nonlinear effects may be neglected.
We choose $\bar n=10$ and $\bar n=100$.
This means that the minimum value of $N$ for which we may solve the
model using our method is of the order of $100$ for $\bar n=10$ and
$1000$ for $\bar n=100$. We follow exactly the same procedure 
as in the
fully stochastic case: We evolve the same initial condition
over $p\times 500$ steps of time, with $p=500$.

The calculations of the velocity of the front
are presented in Tab.~\ref{tab1}. In Fig.~\ref{fig2}, they are 
compared to the results
obtained by solving the exact model. We observe a perfect
agreement for $\bar n=100$, at the level of $0.1\%$. 
For $\bar n=10$, the agreement
is less good: the velocity is overestimated
by about $1\%$ (but this means a $10\%$ mismatch in the difference
$v(\gamma_0)-V$ with the infinite-$N$ velocity). 
Of course, a better agreement than that could not really be expected.
However, part of the mismatch 
may also be related to the fact that
our matching prescription between the dense and dilute regions leads to a
slight overestimate of the number of particles in the lower part of
the dense region, which has indeed the effect of increasing the
front velocity. 
It may well be that for different models which do not have the
requirement that $T$ be a decreasing function of $x$,
like QCD, the agreement would even be better.

The diffusion coefficient $D$ of the front is 
given in Tab.~\ref{tab1} and
displayed in Fig.~\ref{fig3}.
We also plot the theoretical formula~(\ref{theoryD}), to guide the eye.
The numerical data seem to agree well
with the asymptotic theoretical estimate~(\ref{theoryD}), but this must be
accidental since for the considered values of $N$, corrections
are expected to be huge.
The diffusion coefficient is a quantity that is much more difficult 
to measure numerically than the velocity, since it requires more
statistics. Our error bars are still of the order of 5 to 10\%.
We see however that the full stochastic model is very well reproduced,
within errors, by the mixed model with $\bar n=100$.
For $\bar n=10$ instead, we notice that the fluctuations 
seem to be globally underestimated by up to 30\% (see Tab.~\ref{tab1}).
But this should be expected since our method consists in replacing part
of the stochastic evolution by a mean field approximation.

\begin{table}
\begin{tabular}{lllll}
\hline
 & $N$ &stochastic & $\bar n=10$ & $\bar n=100$\\
\hline
\hline
$V$& $10^2$& $1.8937(3)$ & $1.9173(5)$ & \hskip 0.5cm{---} \\
   & $10^3$& $1.9824(2)$ & $1.9910(2)$ &  $1.9838(4)$\\
   & $10^4$& $2.0176(2)$ & $2.0202(3)$ &  $2.0173(3)$\\
   & $10^5$& $2.0348(2)$ & $2.0345(2)$ &  $2.0331(2)$\\
\hline
\hline
$D$ &$10^2$ & $121.2\pm 3.9$ & 
$85.47\pm 4.51$&  \hskip 0.5cm{---} \\
$\times 10^{3}$ &$10^3$ & $41.87 \pm 2.46$& $33.89\pm 2.96$&  $44.19\pm 2.74$\\
 &$10^4$ & $21.99 \pm 0.68$& $16.07\pm 1.21$&  $18.73 \pm 1.36$\\
 &$10^5$ & $11.82 \pm 0.93$& $8.243\pm  0.498$& $10.38 \pm 0.68$\\
\hline
\end{tabular}
\caption{\label{tab1}Numerical results for the one-dimensional toy model.
The velocity $V$ and the diffusion coefficient $D$ of the front are computed
in the full stochastic model and in the two mixed models with factorization scales
$\bar n=10$ and $\bar n=100$ respectively, for four different values of $N$.}
\end{table}


\section{Outlook for QCD}

We have worked on specific toy models of noisy evolution
of the FKPP type. We have proposed that, on an event-by-event basis,
their evolution may be factorized into a linear stochastic part and
a nonlinear deterministic part.
We have shown on a zero-dimensional model that writing down this factorization
leads to a straightforward computation of the average number of particles
at leading order in $t/N$,
without having to go through the resummation of a divergent series like
in more standard approaches \cite{M1994,Salam,SX2005}.

We have conjectured that such a factorization is also valid
for one-dimensional models. In fact, this factorization had already been
implicitly assumed in previous works in statistical physics, essentially
for the purpose of performing numerical calculations for very large values
of $N$ \cite{BD,Moro,EGBM2005}. 
In Ref.~\cite{BDMM2005}, it was even used to
obtain analytical results for the front position, but only
exponentially large values of $N$ were attainable.
We have shown here that this factorization may be extended (in practice numerically) to
lower values of $N$, of the order of $100$ or $1000$.
The only condition is the existence of a mesoscopic scale of particle
numbers $\bar n$ such that $1\ll \bar n\ll N$.
We have tested the factorization in a model that we could also solve
numerically, and have shown that it reproduces quite well
the exact solution even for values of $N$ as low as 100.

Of course, the dense-dilute
factorization would be of no interest for numerical
calculations at low values of $N$ in 
cases in which the complete underlying stochastic model 
were known -- systems
with 100 or 1000 particles may often be simulated quite easily.
However, as we have recalled in the introduction, the exact
formulation of high energy QCD as a stochastic process (if it exists)
has not been found yet.
Instead, we have a probabilistic rule for the evolution 
of the {\it dipole number} $n$ 
valid in the dilute regime
of small amplitudes (see Eq.~(\ref{dipoles})), which has already been
implemented 
numerically by Salam \cite{Salam} (see also \cite{A2004}),
and a mean-field equation for evolving the {\it amplitude} $T$ in the dense
regime (the Balitsky-Kovchegov equation \cite{B,K}, see Eq.~(\ref{BK})).
In our toy model, the latter is equivalent to Eq.~(\ref{evolT}) and the former
is like the evolution rule for our system of particles.
In addition, the relationship between $n$ and $T$ in the dilute regime
is needed: it is provided
by Eq.~(\ref{corresp}) in QCD, and has its equivalent 
(Eq.~(\ref{T-n})) in the toy model
of the previous section.
Consequently, the factorization proposed here should be very well suited
for numerical computations of QCD scattering amplitudes at high energy: 
One should combine 
Salam's dipole Monte-Carlo and a solution of the Balitsky-Kovchegov equation,
in a way similar to what we have done here for the one-dimensional toy model.
The details will be worked out in another publication \cite{IMMS}.

Extrapolating the results of 
our toy model study, we can expect to get reliable 
results for models in which
the typical allowed number of particles $N$ is larger than $100$. This corresponds
to $\alpha_s<0.1$ in QCD, which is not far from the experimentally accessible
window. Hence the factorization procedure outlined here may be a good
starting
point for a realistic numerical investigation
of QCD amplitudes near the unitarity limit. Anyway, it is probably
the best one could achieve without finding a realization of 
nonlinear saturation effects in
the dipole model.
Needless to say, the toy model that we have studied here
was tuned to make our factorization procedure as easy as possible to
handle numerically.
The many complications of QCD will make the implementation 
of this factorization a challenging issue.

At a more theoretical level, what our study suggests is that
universal features show up already for quite low values of $N$.
This leaves us with the hope that
the asymptotic analytical calculations of Ref.~\cite{BDMM2005}
could be extended to a wider range in $N$.

Finally, we wish to comment on a recent proposal on how to
address numerically the problem of QCD evolution at very high energies beyond
the mean-field BK limit.
In Ref.~\cite{KL}, 
the idea that fluctuations could be obtained
by solving a classical equation and then averaging over an ensemble of
initial conditions was suggested,
and was implemented numerically subsequently in Ref.~\cite{AM}.
We note that the philosophy of this proposal is orthogonal to the one
developed here: In our view, fluctuations in the saturation scale are
due to intrinsic noise related to the discreteness of the number of
partons, as is natural in reaction-diffusion processes.
We do not think that the approach of Refs.~\cite{KL,AM} would 
work for reaction-diffusion at very large times (i.e. asymptotic energies):
For example, a universal distinctive feature of such processes
is that the dispersion of the
front positions (i.e. saturation scales)
between different realizations (i.e. events)
scales like $\sqrt{Y}$, which cannot be reproduced in the approach
of Refs.~\cite{KL,AM}.
Since our numerical method relies on the conjecture that QCD 
evolution is in the same universality 
class as reaction-diffusion processes, 
the two approaches would
not lead to the same results for QCD observables
(see Ref.~\cite{AM}).
However, since the
statement that there is a correspondence between 
QCD and reaction-diffusion
is still a conjecture, one has to keep open to
other possibilities.\\


\begin{acknowledgments}
We thank Prof. A. H. Mueller for
many decisive discussions, and B.-W. Xiao for
explaining the details of
his resolution of the zero-dimensional model presented
in Ref.~\cite{SX2005}.
We are grateful to Prof. A. H. Mueller and Prof. E. J. Weinberg
for welcome at Columbia University and support
at the time when this work was initiated.
We also thank Dr. G. Soyez for a discussion about the technical
difficulties of implementing the factorization in QCD,
and Dr. K. Golec-Biernat and Dr. L. Motyka for encouraging discussions.
\end{acknowledgments}


\end{document}